# *Thermodynamic Integration Methods, Infinite Swapping and the Calculation of Generalized Averages*

J. D. Doll
Department of Chemistry
Brown University
Providence, RI 02912

and

P. Dupuis and P. Nyquist
Division of Applied Mathematics
Brown University
Providence, RI 02912

## *Abstract*

In the present paper we examine the risk-sensitive and sampling issues associated with the problem of calculating generalized averages. By combining thermodynamic integration and Stationary Phase Monte Carlo techniques, we develop an approach for such problems and explore its utility for a prototypical class of applications.

***I. Introduction:*** Monte Carlo methods[1] are a valuable tool for the study of many-dimensional problems in a variety of disciplines. By providing a general means of investigating the properties of well-defined, physically realistic models without resorting to untestable numerical approximations, they are an essential element in obtaining the insight necessary for the construction of valid conceptual models of complex phenomena.

An important use of Monte Carlo methods is providing numerical estimates of averages of the type that arise naturally in a variety of classical[2] and quantum statistical-mechanical contexts.[3,4] In the present work we wish to consider the computation of generalized averages of the form

$$<e^{bx}>_S = \frac{\int e^{-S(x,\alpha)} e^{bx} dx}{\int e^{-S(x,\alpha)} dx} ,$$

(1.1)

where S and b are both potentially complex. For convenience we utilize a pseudo one-dimensional notation in the following with the understanding that multidimensional generalizations of all results are readily obtained. In Eq. (1.1) x and $\alpha$ represent the coordinate(s) and system parameters of the problem, respectively. Real valued moment generating functions of this type are common in the calculation of equilibrium thermodynamic properties.[1,2] In quantum dynamical applications, on the other hand, S becomes complex.[5-7] Averages of the type in Eq. (1.1) provide a useful class of prototypes for both classes of problems in that they represent a general demonstration of the ability to compute analogous averages of any functions that can be written in Fourier or Laplace form.

In confronting generalized averages of the type in Eq. (1.1), there are a number of core issues. Chief among them are the choice of an appropriate sampling density and the selection/design of sampling methods to assure all regions of importance for that density are properly included in the final average. In the present work we combine Stationary Phase Monte Carlo (SPMC)[5-7] and Infinite Swapping (INS) techniques[8-10] to accomplish these twin tasks. The INS computational ensemble in the present developments is based on a spatial rather than a thermal control parameter. In addition to the utilization of improved sampling methods, we find that a reformulation of the underlying problem using Kirkwood style thermodynamic integration techniques[2,11] proves advantageous.

The remainder of this paper is organized as follows. In Section II we examine the risk sensitive nature of moment generating functions of the type in Eq. (1.1) and discuss the advantages of a Kirkwood approach for their evaluation. Methods for implementing the Kirkwood approach for complex generalizations of Eq. (1.1) are presented in Section III, and illustrative examples are presented and discussed in Section IV.

***II. Risk Sensitivity and the Kirkwood Formulation:***  In considering the evaluation of averages of the type in Eq. (1.1) it is useful to note that there are a number of possible approaches. One is the direct application of Monte Carlo methods.[1] Specifically, if $S(x,\alpha)$ is real, a natural route is to replace the continuous average in Eq. (1.1) with the discrete average of the integrand, $\exp(bx)$, over a finite set of points obtained from a Monte Carlo sampling of the density, $\exp(-S(x,\alpha))$. If $S(x,\alpha)$ is not real, a case examined in greater detail in the following Section, the choice of an appropriate density for use as an importance function becomes more subtle. In either situation rather than approaching Eq. (1.1) directly it proves useful first to restructure it.

Using techniques familiar from "thermodynamic integration" methods[2,11] the average we seek can be recast exactly as

$$<e^{bx}>_S = \exp\left(\int_0^b <x>_\lambda \, d\lambda\right),$$

(2.1)

where

$$<x>_\lambda = \frac{\int e^{-S_\lambda(x,\alpha)} x \, dx}{\int e^{-S_\lambda(x,\alpha)} dx},$$

(2.2)

and where

$$S_\lambda(x,\alpha) = S(x,\alpha) - \lambda x .$$

(2.3)

For brevity, we refer to expressions such as Eq. (2.1) in what follows as the "Kirkwood" form of the original average, Eq. (1.1).

While the Kirkwood and original forms of the average are equally valid, they can differ significantly in their sensitivities to statistical noise. Anticipating an ultimate evaluation by stochastic means, this difference in sensitivity is potentially an important practical matter. The simple case where b is real and S is a quadratic serves to illustrate this point. Assuming S to be of the form

$$S(x,x_0) = \frac{1}{2}(x - x_0)^2 ,$$

(2.4)

where $x_0$ is a real constant, the average specified by Eq. (1.1) is given analytically by

$$\left\langle e^{bx} \right\rangle_{Exact} = e^{bx_0 + \frac{b^2}{2}} ,$$

(2.5)

while the standard deviation of exp(bx) with respect to exp(-S(x,$x_0$)) is given by

$$\sigma = \left[ <e^{2bx}>_S - <e^{bx}>_S^2 \right]^{1/2} = e^{bx_0 + \frac{b^2}{2}} (e^{b^2} - 1)^{1/2} .$$

(2.6)

From Eqs. (2.5) and (2.6) we see that the error associated with an N-point direct Monte Carlo (DMC) estimate of Eq. (1.1) scales poorly with respect to the parameter b. In particular, assuming N independent Monte Carlo sample points drawn from the density associated with exp(-S(x,$x_0$)), the DMC estimate of the moment generating function, $\left\langle e^{bx} \right\rangle_{DMC}$, is given by the exact value plus a random variable whose standard deviation is $\sigma/N^{1/2}$. In other words, the ratio of the DMC estimate of the moment generating function to its exact value scales as

$$\frac{\langle e^{bx}\rangle_{DMC}}{\langle e^{bx}\rangle_{Exact}} = 1 + \xi_{DMC},$$

(2.7)

where $\xi_{DMC}$ is a random variable whose standard deviation is given by $(e^{b^2}-1)^{1/2}/\sqrt{N}$. This result makes the direct approach unworkable except for small values of b. Such extreme sensitivity to noise is a general characteristic of "risk-sensitive" problems[12] in which the variance of the integrand in question is dominated by regions of the underlying integration that are in the tails of the importance function.

A similar analysis of the errors in the thermodynamic integration or Kirkwood (K) method yields a markedly different outcome. Assuming that the variance in each of the $<x>_\lambda$ terms in Eq. (2.1) is independent of λ and that the λ-quadrature is based on a fixed grid size (i.e. the number of quadrature points required for the numerical λ-integration increases linearly with the size of the integration domain, b), it is straightforward to show that the ratio analogous to that in Eq. (2.7) for the Kirkwood approach in the large N limit is given by

$$\frac{\langle e^{bx}\rangle_{K}}{\langle e^{bx}\rangle_{Exact}} = 1 + \xi_{K},$$

(2.8)

where $\xi_K$ is a random variable whose standard deviation is given by $\sqrt{b/N}$. The computational moral to this story is that the Kirkwood and direct approaches can differ significantly with respect to their sensitivities to Monte Carlo noise. This difference offers a potential means for avoiding/dealing with risk-sensitive issues that arise in the application of Monte Carlo techniques. It should also be noted that the Kirkwood approach avoids the explicit calculation of partition function ratios or their analogs.

***III. Methods for Complex Averages:*** A number of practical issues arise in situations where S(x,α) in Eq. (1.1) is complex. A major one is the choice of an importance function. A tempting choice is the modulus, |exp(-S)|. However, the (potentially) highly

oscillatory nature of the integrand means that the important regions of the integrand are no longer dictated exclusively by |exp(-S)|, but by a competition between that modulus and the stationary phase regions of the problem.

Stationary Phase Monte Carlo (SPMC) techniques, described in detail elsewhere,[5-7] have been developed for dealing with complex averages. Briefly summarized, these approaches are based on the observation that there exists a group of transformations of integrands that leave the value of the associated integrals unchanged. Specifically, for integrals over an infinite or periodic domain and which converge sufficiently rapidly such that orders of integration can be interchanged, integrals of a function and of its convolution with an arbitrary probability density are equal. That is, given a function f(x) and a normalized probability density $P_\varepsilon(y)$, we have

$$\int f(x)\,dx = \int <f(x)>_\varepsilon dx,$$

(3.1)

where

$$<f(x)>_\varepsilon = \int P_\varepsilon(y) f(x+y)\,dy.$$

(3.2)

Although the left and right hand sides of (3.1) are equal, the corresponding integrands generally differ. The "pre-averaging" process in Eq. (3.2) damps the integrand's oscillations on a controllable length scale. Applying this idea to the average in Eq. (2.2), we have

$$<x>_\lambda = \frac{\int \left\langle e^{-S_\lambda(x,\alpha)} x \right\rangle_\varepsilon dx}{\int \left\langle e^{-S_\lambda(x,\alpha)} \right\rangle_\varepsilon dx}.$$

(3.3)

The result in Eq. (3.3) is formally independent of the parameter(s) $\varepsilon$. In practice, the variation of x relative to that of exp(-S(x,$\alpha$)) on the length scale $\varepsilon$ is often small. Under such conditions Eq. (3.3) is well approximated by the computationally more convenient expression

$$<x>_\lambda = \frac{\int \left\langle e^{-S_\lambda(x,\alpha)} \right\rangle_\varepsilon x\, dx}{\int \left\langle e^{-S_\lambda(x,\alpha)} \right\rangle_\varepsilon dx}.$$

(3.4)

Unlike |exp(-S)| itself, if the ε-length scale is properly chosen, $W_\varepsilon(x)$, defined as

$$W_\varepsilon(x) = \left| \left\langle e^{-S(x,\alpha)} \right\rangle_\varepsilon \right|,$$

(3.5)

does provide a suitable importance function for the evaluation of the average in Eq. (3.4). In practical terms we require that ε be chosen small enough that the approximation of replacing Eq. (3.3) by (3.4) is valid, but large enough that the irrelevant, non-stationary phase regions are suppressed and the important regions emphasized. Issues related to the choice of ε have been discussed previously[13] and will be examined in greater detail in the following Section.

Consistent with the working assumption that the length scale ε is small, we approximate the ε-average in $\left\langle e^{-S(x,\alpha)} \right\rangle_\varepsilon$ using gradient methods. Assuming a Gaussian form for $P_\varepsilon(y)$, through second-order the gradient approach gives

$$\left\langle e^{-S(x,\alpha)} \right\rangle_\varepsilon = \frac{\exp\left\{-S(x,\alpha) + \frac{1}{2}(\varepsilon S'(x,\alpha))^2 / (1+\varepsilon^2 S''(x,\alpha))\right\}}{(1+\varepsilon^2 S''(x,\alpha))^{1/2}},$$

(3.6)

where S' and S'' denote the first and second derivatives of S, respectively. Analogous first-order and multidimensional approximations are easily derived. The second-order gradient approximation to the SPMC importance function is given by the modulus of Eq. (3.6).

Monte Carlo applications of the type under discussion frequently involve sparse sampling issues. When the probability density that underlies the average in question is composed of isolated or weakly connected regions, special care must be exercised to assure that all regions of importance are properly included. Failure of the sampling method to provide a proper accounting is, in practice, both computationally destructive and difficult to detect. Such difficulties, present in conventional real-valued forms of Eq. (1.1), become even more problematic in analogous complex averages where the potentially highly oscillatory nature of the integrand plays a key role. The SPMC approach is, in essence, the exchange of a problem involving severe phase oscillations for one of sparse sampling.

A variety of techniques have been developed for dealing with the general sparse sampling problem. One approach, parallel tempering,[14-16] utilizes a computational ensemble composed of the product of densities for a set of control parameters (typically the temperature). Rather than studying the various ensemble members individually, parallel tempering studies the entire ensemble in unison. Ordinary random walk displacements are augmented with trial moves based on attempted swaps of configurations between the different data streams. By demanding that detailed balance be preserved for such swaps, the resulting approach provides a practical means for using information from the more highly-connected members of the ensemble to improve the efficiency of sampling for the more weakly connected densities.

The recently developed Infinite Swapping (INS) approach[8-10] is a sparse sampling strategy based on a large deviation analysis of parallel tempering. It represents the extreme limit of parallel tempering in which swaps involving all possible temperatures are attempted at an infinitely rapid rate, a limit the large deviation analysis proves to be optimal. Operationally, the method utilizes a probability density composed of a *symmetrized* sum of parallel tempering like product densities, a form that is more highly connected than the original. The INS approach represents the conscious use of symmetry as a tool for dealing with the sparse sampling problem. Practical methods for implementing the approach for arbitrary sized ensembles have been developed and discussed in detail elsewhere.[8-10]

We propose an approach to the construction of averages of the type in Eq. (2.1) that consists of three elements:

- A Kirkwood-like formulation of the problem to deal with its risk sensitive aspects;
- SPMC methods to suppress phase oscillations and to produce a suitable importance function (Eq. (3.5)); and
- INS techniques to treat the sparse sampling issues arising from use of the SPMC approach.

In conventional parallel tempering simulations the system temperature is typically utilized as the control parameter for the creation of the computational ensemble. The various data streams within such simulations thus produce estimates of thermodynamic properties for the various temperatures within the ensemble. In the present work, on the other hand, the control parameter for the INS ensemble is the SPMC length scale, $\varepsilon$. Recalling that the overall SPMC results are independent of the choice of this length scale (c.f. Eq. (3.1)), we see that the different data streams in the present approach are estimates of the *same* computational object. Because they correspond to different SPMC length scales, however, the *quality* of these estimates will generally differ.

*IV. Results and Discussion:* In this Section we illustrate the current approach with an application to a model average of the form

$$\phi(\eta) \;=\; <e^{i\eta x}>_S \;=\; \frac{\int e^{-S(x)} e^{i\eta x} dx}{\int e^{-S(x)} dx},$$

(4.1)

where S(x) is a complex quantity specified by

$$S(x) = \frac{1}{2}\left(\frac{x-x_0}{\sigma}\right)^2 - i\left(\frac{x^3}{3}\right).$$

(4.2)

To streamline the notation in Eq. (4.2) and in the following discussion the explicit $x_0$ and $\sigma$ labels in the expression for S(x) will be omitted. Simple enough that key aspects of the method can be readily investigated, the present model is sufficiently complex to reflect the general computational challenges involved.

The nature of the underlying average in Eq. (4.1) changes character as a function of the parameters $\eta$ and $\sigma$. In the small $\eta$ limit, $|\exp(-S)|$ covers the important regions of the problem making conventional Monte Carlo methods generally applicable. In the limit that $\eta$ is large and negative, however, the phase oscillations for the integrand in the numerator of Eq. (4.1) become severe and the stationary phase regions at $\pm(-\eta)^{1/2}$ play a dominant role. Depending on the value of $\sigma$, these stationary phase regions may or may not fall within the natural range of $|\exp(-S)|$. In any case in a conventional Monte Carlo approach the irrelevant regions of the problem would be established in an inefficient, after-the-fact manner through numerical cancellation involving poorly placed Monte Carlo points. In contrast, if the SPMC length scale is properly chosen, the importance function $|<\exp(-S)>_\varepsilon|$ is concentrated in the regions that dominate the final result and the inefficient numerical cancellation issue is avoided.

The Kirkwood form of Eq. (4.1) is

$$\phi(\eta) = \exp\left(i\int_0^\eta <x>_\lambda d\lambda\right)$$

(4.3)

where

$$<x>_\lambda = \frac{\int e^{-S_\lambda(x)} x dx}{\int e^{-S_\lambda(x)} dx},$$

(4.4)

and where

$$S_\lambda(x) = S(x) - i\lambda x.$$

(4.5)

To produce an estimate of the original computational objective in the Kirkwood approach the essential numerical tasks are to evaluate $<x>_\lambda$ on a grid of $\lambda$-values and to perform the associated one-dimensional $\lambda$–integration. To calculate the necessary $<x>_\lambda$ values, we rewrite Eq. (4.4) using the SPMC methods of Section III as

$$<x>_\lambda = \frac{\int \left\langle e^{-S_\lambda(x)} \right\rangle_\varepsilon x\, dx}{\int \left\langle e^{-S_\lambda(x)} \right\rangle_\varepsilon dx}.$$

(4.6)

INS techniques can then be used to evaluate $<x>_\lambda$ as a function of $\lambda$ for each of the $\varepsilon$-values in the ensemble. Once the necessary $<x>_\lambda$ values are prepared, conventional numerical quadrature techniques can be used to perform the $\lambda$-integration in Eq. (4.3). Unless otherwise noted all numerical results presented in the present studies utilize:

- second-order gradient approximations (Eq. (3.6)) for the necessary SPMC averages,
- a 5-member INS ensemble based on a range of $\varepsilon$ values ($\varepsilon$ = (0.00,0.05,0.10,0.20,0.40)) chosen by methods outlined below,
- Metropolis single-variable techniques in combination with the heat bath method outlined previously[10] to perform the necessary sampling, and
- trapezoidal quadrature to perform the one-dimensional $\lambda$-integration in Eq. (4.3).

We begin by first investigating the $\varepsilon$-independence of the results of Eq. (4.6). From the discussion in Section III we know that the results of Eq. (3.3) are formally independent of the choice of the $\varepsilon$ parameter. Table I examines the extent to which this is also true of the approximate result in Eq. (4.6) for the set of five $\varepsilon$-values and system parameters used in the present studies. Shown in Table I are the numerical values of $<x>_\lambda$ for the present model as a function of $\varepsilon$ for two representative, large negative values of $\lambda$. All results in Table I are computed for $\sigma = 1$ and $x_0 = 0.5$ using Mathematica to perform the necessary integrations. The simplicity of the present model problem permits the use of such conventional methods to provide an unambiguous test of the level of $\varepsilon$-independence of Eq. (4.6). More generally, the presence or absence of such $\varepsilon$–independence will in practice be signaled by the internal consistency of the calculated results for the various

INS ensemble members. We see from Table I that the $\langle x \rangle_\lambda$ results display only a very weak dependence on $\varepsilon$ over the range studied thus justifying the use of Eq. (4.6).

Table I

| $\varepsilon$ | $\langle x \rangle$ ($\lambda = -16$) | $\langle x \rangle$ ($\lambda = -25$) |
|---|---|---|
| 0.00 | 3.822 - 0.463 i | 4.926 - 0.412 i |
| 0.05 | 3.822 - 0.463 i | 4.926 - 0.412 i |
| 0.10 | 3.822 - 0.463 i | 4.926 - 0.412 i |
| 0.20 | 3.822 - 0.461 i | 4.926 - 0.411 i |
| 0.40 | 3.894 - 0.442 i | 4.975 - 0.435 i |

Shown in Fig. (1) are the (normalized) importance functions, $W_\varepsilon(x)$, for the $\varepsilon$ values of Table I obtained from the modulus of the corresponding second-order result (Eq. (3.6)) for various values of $x_0$, $\lambda$ and $\sigma$. The color/$\varepsilon$-assignments involved are listed in the figure caption. To facilitate the comparison of the various results all densities in Fig. (1) are normalized to unity.

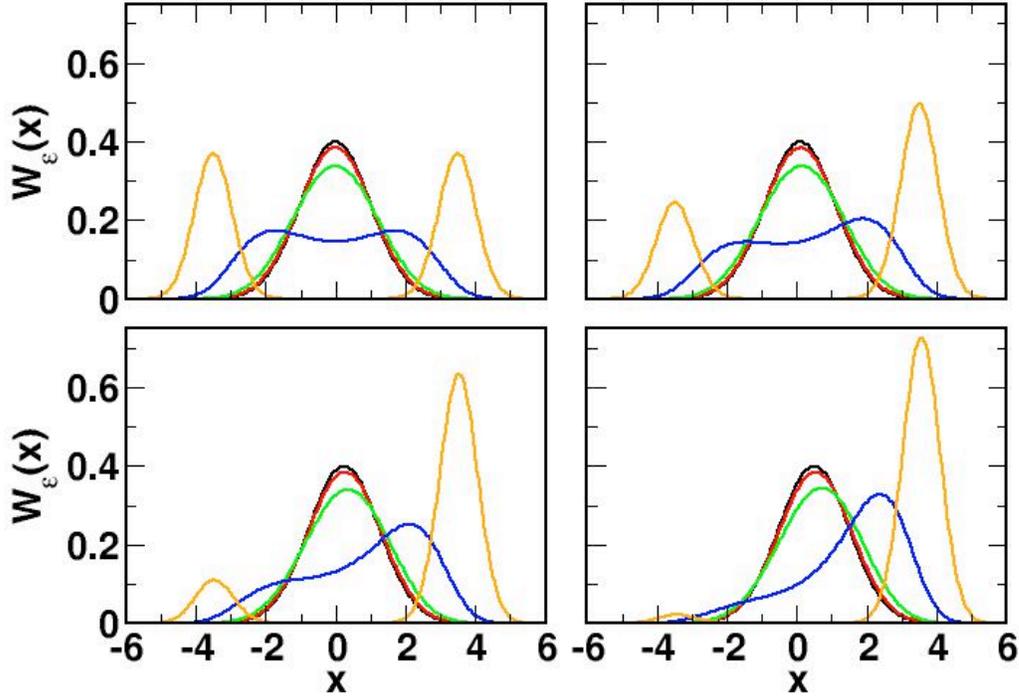

Fig. 1: Normalized $W_\varepsilon(x)$ plots for $\lambda = -16$, and $\varepsilon = 0.00$ (black), 0.05 (red), 0.10 (green), 0.20 (blue) and 0.40 (orange). The $(\sigma, x_0)$ values are (1,0.0) upper left, (1,0.50) upper right, (2,0.0) lower left and (2,0.5) lower right.

The changing character of the importance functions with SPMC length scale is evident in the results of Fig. (1). For smaller $\varepsilon$, $W_\varepsilon(x)$ results reflect the real portions of $S(x)$ and are thus essentially unimodal Gaussians centered on the corresponding values of $x_0$. As $\varepsilon$ increases, this initial unimodal density tends to increase in width and then ultimately to develop a structure that reflects the underlying stationary phase regions of the problem. The sharpness of the resulting stationary phase feature(s) varies with $\varepsilon$, being most highly focused when the $\varepsilon$ length scale matches the natural width(s) of those region(s).

The $\varepsilon$-dependence of the SPMC density is conveniently summarized by the associated information entropy. Shown in Fig. (2), for example, are plots of the information entropy for the densities of the system described in Fig. (1a) for $\lambda = -16$ and $-25$ as a function of the SPMC length scale, $\varepsilon$. The increases in entropy visible in the small $\varepsilon$ regions of Fig. (2) correspond to the broadening of the initial unimodal Gaussian densities centered at $x_0$. As $\varepsilon$ continues to increase, the information entropies peak, go through minima, and ultimately increase as the initial unimodal densities first split, sharpen, and then broaden. The peak in the information entropy as a function of $\varepsilon$ thus serves as a rough indicator of the $\varepsilon$ value for which the SPMC importance function begins to reflect qualitatively the inherent stationary phase character of the problem, roughly 0.2 for the systems in Fig. (2). The minimum in the information entropy, on the other hand, provides a practical guide for the the $\varepsilon$ value that produces the maximally compressed SPMC density, roughly 0.4 for the systems in Fig. (2). Such considerations form the basis for the selection of the INS computational ensemble for the present example. In general applications, information entropy *differences* rather than the *absolute* entropies provide a more readily computed basis for such decisions.

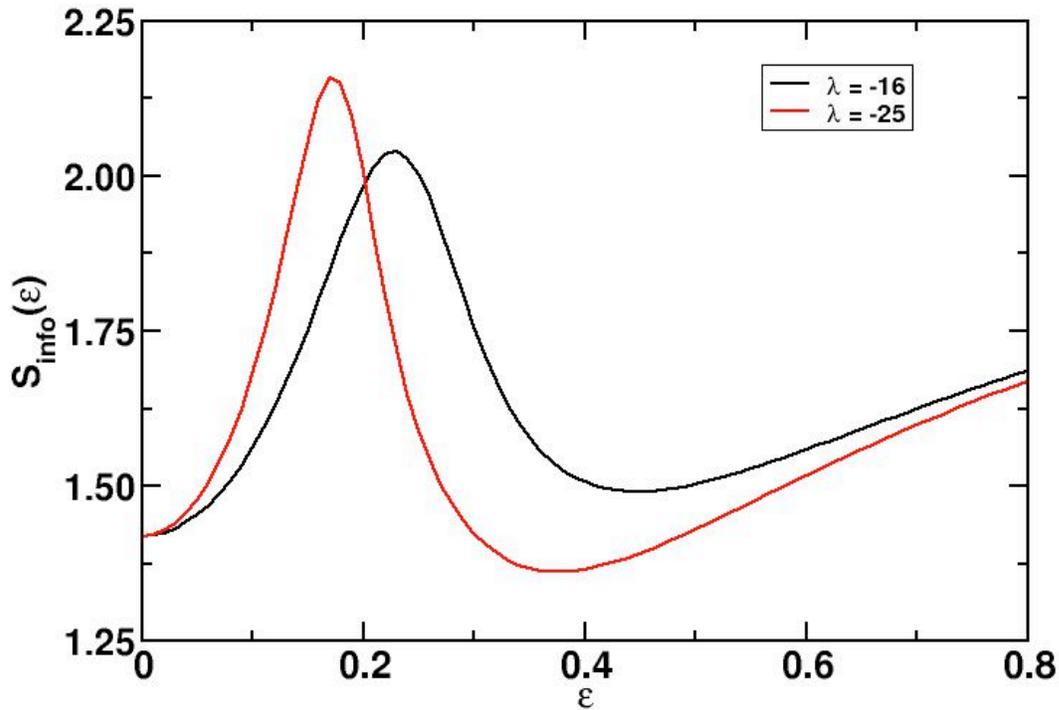

Fig. 2  Plots of the information entropy, $S_{info}(\varepsilon)$, as a function of the SPMC length scale $\varepsilon$ for the parameters of Fig. (1a) for $\lambda = -16$ (black) and $\lambda = -25$ (red).

Shown in Fig. (3) are the real and imaginary parts of $\langle x \rangle_\lambda$ as a function of $\lambda$ computed from Eq. (4.6) using the present INS/SPMC approach. The $\langle x \rangle_\lambda$ results shown are those for $\sigma = 1$, $\varepsilon = 0.40$ for two different choices of $x_0$ obtained using $10^6$ single particle Monte Carlo moves for each of a discrete grid of $\lambda$-values (grid spacing = 0.05). As can be seen in the large $\varepsilon$-results of Fig. (1), the stationary phase regions that dominate the present averages for large negative $\lambda$-values are isolated and represent a small fraction of the total integration volume. The resolution and detail of the results in Fig. (3) indicate that the INS approach is effective in dealing with the rare-event sampling issues involved.

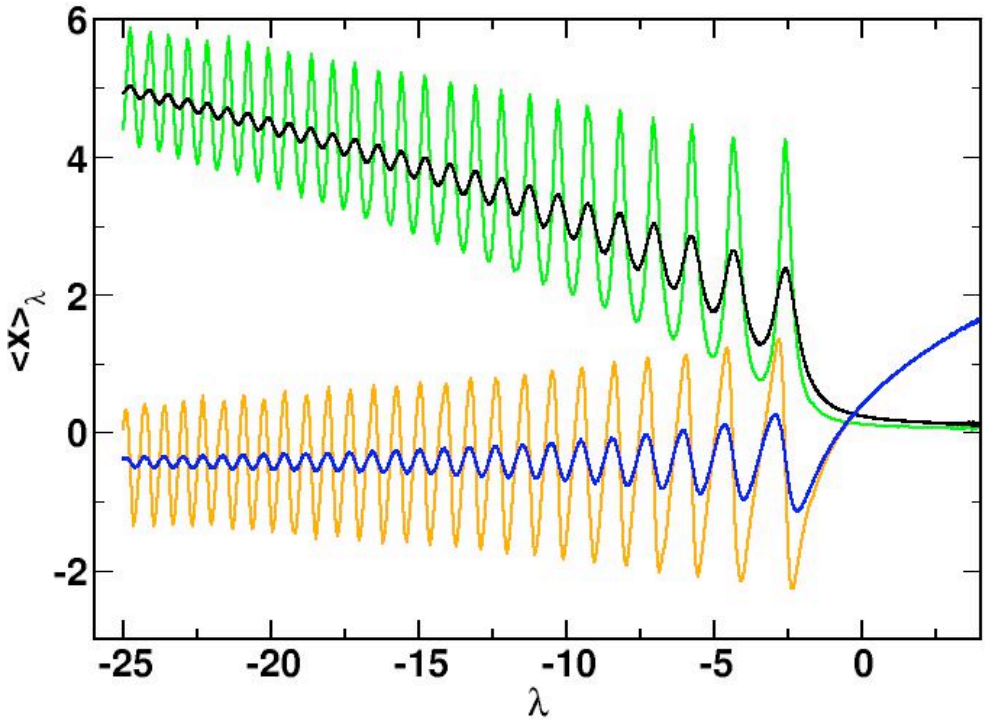

Fig. 3: $\langle x \rangle_\lambda$ for Gauss/Airy model as a function of $\lambda$. Results computed using the SPMC/INS approach described in greater detail in the text. Green and orange plots correspond to real and imaginary parts for the choice of $\sigma = 1$, $x_0 = 0.25$ while the black and blue traces are the corresponding results for $\sigma = 1$, $x_0 = 0.50$. All results obtained using the SPMC parameter $\varepsilon = 0.40$ using $10^6$ total Monte Carlo points (100 loops of $10^4$ points).

To validate the results for the chosen model and to understand better the performance of the present computational approach, it is useful to examine selected sequences of results. Shown in Fig. (4) are a number of $<x>_\lambda$ results obtained using various system parameters and numbers of Monte Carlo points. For simplicity we display only the real portions of $<x>_\lambda$. The behavior of the analogous imaginary quantities is qualitatively similar in all cases. We consider first the $<x>_\lambda$ values for a *fixed* number of Monte Carlo points for *varying* $\varepsilon$-values and then examine analogous $<x>_\lambda$ results for *varying* numbers of Monte Carlo points for a *fixed* value of $\varepsilon$. The blue, red, and black curves in Fig. (4) show the real portions of $<x>_\lambda$ values obtained for $x_0 = 0.5$ and $\sigma = 1$ for three of the five ensemble $\varepsilon$ values, $\varepsilon = (0.10, 0.20, 0.40)$, respectively, using $10^6$ Monte Carlo points. The associated results for $\varepsilon = 0.05$ and $0.00$ (not shown) are qualitatively similar to those of $\varepsilon = 0.10$. Although "noisier," the $<x>_\lambda$ results for $\varepsilon = 0.20$ (red curve) are in basic agreement for those for $\varepsilon = 0.40$ (black curve-obscured by red curve) over the entire $\lambda$-range shown in Fig. (4). The level of the agreement between the $\varepsilon = 0.20$ and $0.40$ results is shown in greater detail in Fig. (4a). The $<x>_\lambda$ results for $\varepsilon = 0.10$ (blue curve) in Fig. (4), on the other hand, agree with those of the larger $\varepsilon$ values for the smaller $\lambda$-range (albeit with greater noise), but exhibit systematic errors for large negative $\lambda$-values. At first glance these systematic errors for large negative $\lambda$-values seem inconsistent with the results of Table I. It is important to note, however, that the results in Table I utilize high-precision, direct quadrature while those in Fig. (4) are Monte Carlo estimates based on a *fixed* number of points ($10^6$). From Fig. (1) we see that the importance function for $\varepsilon = 0.10$ poorly reflects the relevant stationary phase regions. The $\varepsilon = 0.10$ importance function has appreciable density in the non-stationary phase regions, regions whose unimportance must then be retroactively established by the use of more sample points. Thus, while Table I tells us that the calculated $<x>_\lambda$ values are, *in principle*, independent of the choice of $\varepsilon$, the results of Fig. (4) tell us that a statistical estimate made using a *fixed* number Monte Carlo points is dependent upon the quality of the associated importance function. In the present case, $10^6$ Monte Carlo points are insufficient to produce all the phase cancellations necessary to overcome the qualitatively incorrect importance function associated with $\varepsilon = 0.10$. This conclusion is reinforced by the brown, orange, blue and green curves of Fig. (4). These curves denote the real portions of the $<x>_\lambda$ results computed for $\varepsilon = 0.10$ for $10^4$, $10^5$, $10^6$, and $10^7$ Monte Carlo points, respectively. We see that the onset of systematic errors in these $<x>_\lambda$ results correlates with the number of Monte Carlo points used in the corresponding simulation. As more points are used, more of the deficiencies of the underlying importance functions are overcome and the $<x>_\lambda$ values are computed reliably for larger negative $\lambda$-values. As illustrated by the black and green curves in Fig. (4), however, improving the underlying importance function is generally a more efficient option than the brute-force approach. Finally, it is important to note that while they individually may have computational shortcomings, the small $\varepsilon$-values of the ensemble actually play a critical role in the INS approach. In particular, they provide the "connective tissue" that bridges the otherwise sparse densities associated with other control parameters. In general, the level of agreement between results computed for different ensemble control parameters serves as a practical internal quality control indicator for the overall simulation.

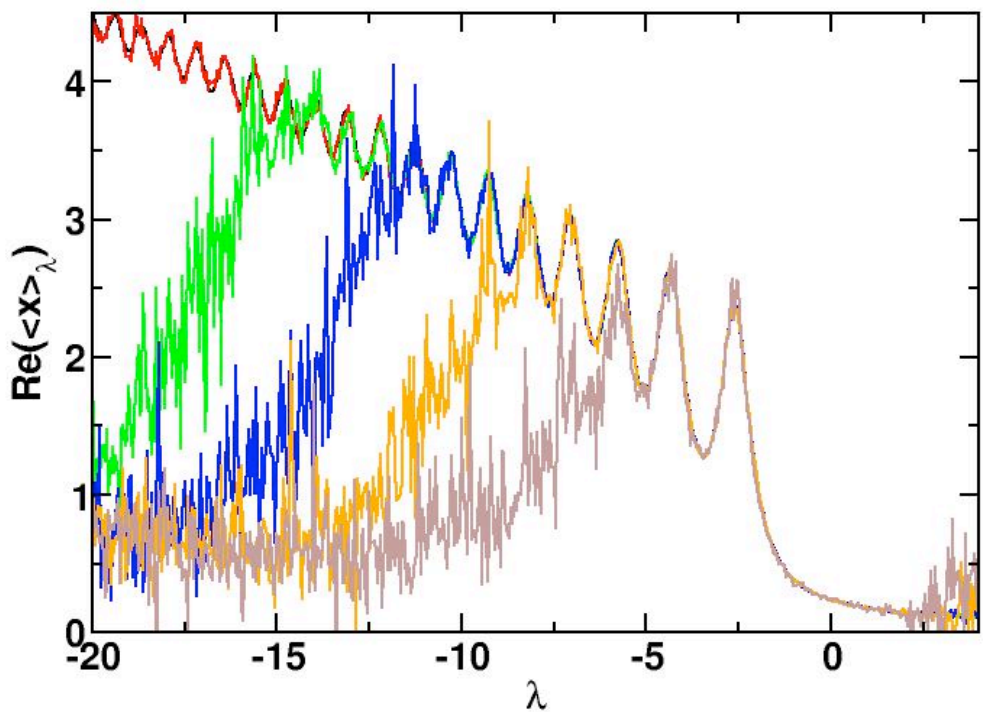

Fig. 4: Additional detail of Re($<x>_\lambda$) for Gauss/Airy model of Fig. (3) ($x_0 = 0.50$) as a function of $\lambda$ for various ε-values. Results for ε = 0.40 (black) and 0.20 (red) are computed using 100 loops of $10^4$ points. Results are shown for ε = 0.10 computed using 100 loops of $10^5$ points (green), $10^4$ points (blue), $10^3$ points (orange), and $10^2$ points (brown). σ = 1 for all results

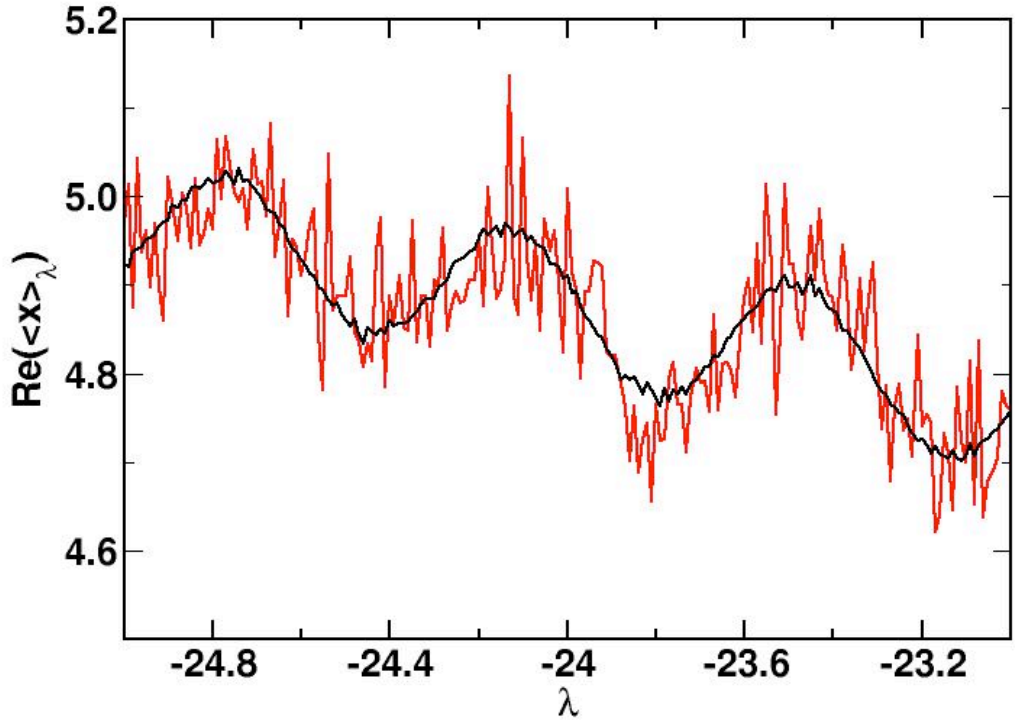

Fig 4a: Blowup of large negative $\lambda$ region of Fig. (4). Shown are the real parts of $\langle x \rangle_\lambda$ as a function of $\lambda$ for $x_0 = 0.50$ obtained using $\varepsilon = 0.40$ (black) and $\varepsilon = 0.20$ (red). Both simulations utilized 100 loops of $10^4$ MC points and $\sigma = 1$.

Figures (5) and (6) show the real and imaginary parts of $\phi(\eta)$ computed from Eq. (4.3) using the $\varepsilon = 0.40$ $\langle x \rangle_\lambda$ results of the type shown in Fig. (3). These results illustrate the variation of the $\phi(\eta)$ results for different $x_0$ values (0.25 = black, 0.50 = red, 1.00 = green) for a fixed value of $\sigma$ (1.00). In general, the results of the type in Figs. (5) and (6) are accurate over the $\eta$-range for which the corresponding $\langle x \rangle_\lambda$ results are $\varepsilon$-independent.

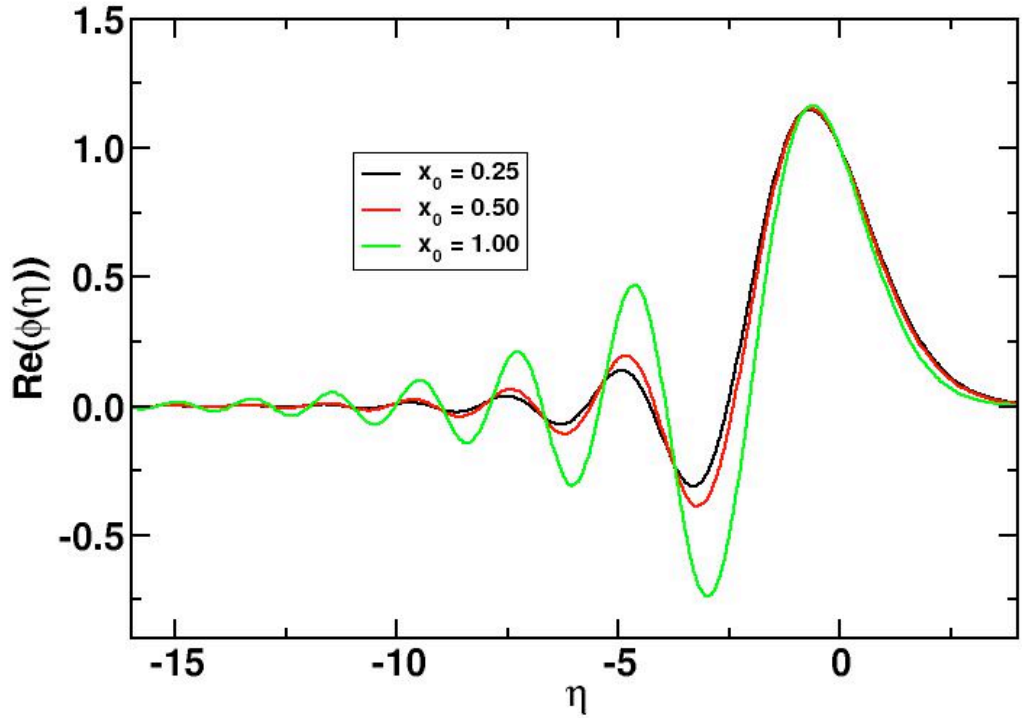

Fig. 5: Plots of the real parts of $\phi(\eta)$ for Gauss/Airy model computed via the Kirkwood approach using $<x>_\lambda$ information from SPMC/INS calculations discussed in the text. Results shown correspond to $\varepsilon = 0.40$ and were computed using 100 loops of $10^4$ points for $x_0 = 0.25$ (black), 0.50 (red) and 1.00 (green). $\sigma = 1$ for all results.

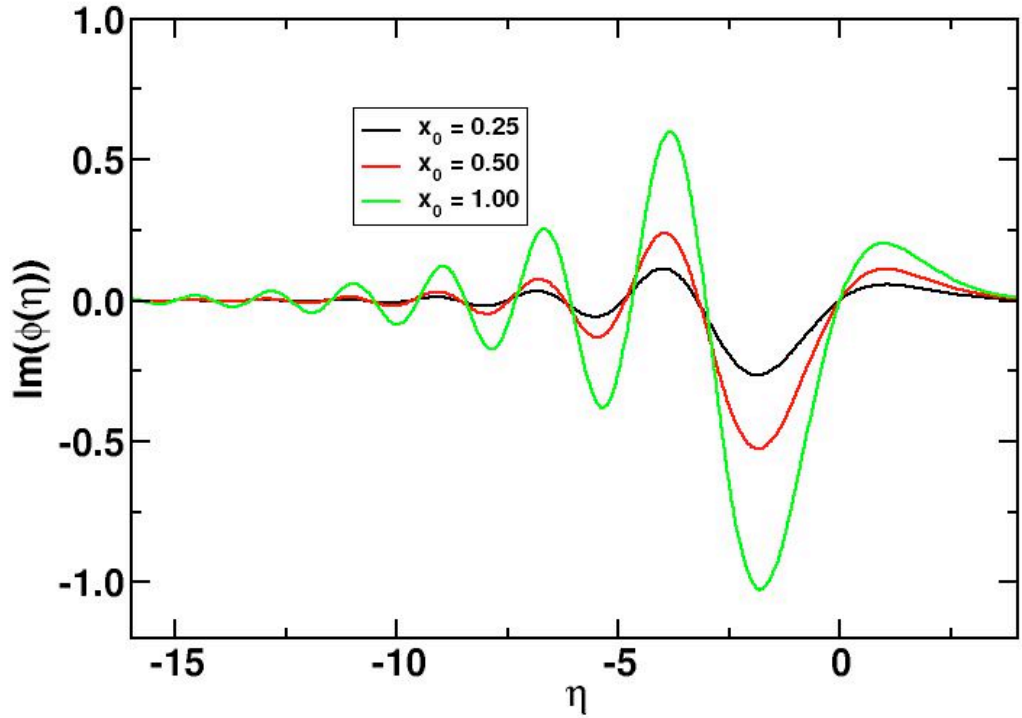

Fig. 6: As in Fig. (5), but for the imaginary parts of $\phi(\eta)$.

Figure (7) documents the ability of the present approach to compute $\phi(\eta)$ accurately for large, negative $\eta$-values, regions hard to treat with direct Monte Carlo methods. For the choice of $x_0 = 0.50$ and $\varepsilon = 0.40$, the red curve in Fig. (7) shows $\text{Re}(\phi(\eta))$ obtained using the present approach while the black curve shows the corresponding results obtained using direct Monte Carlo methods with the same number of points ($10^6$). The corresponding results for $\text{Im}(\phi(\eta))$ are of similar quality.

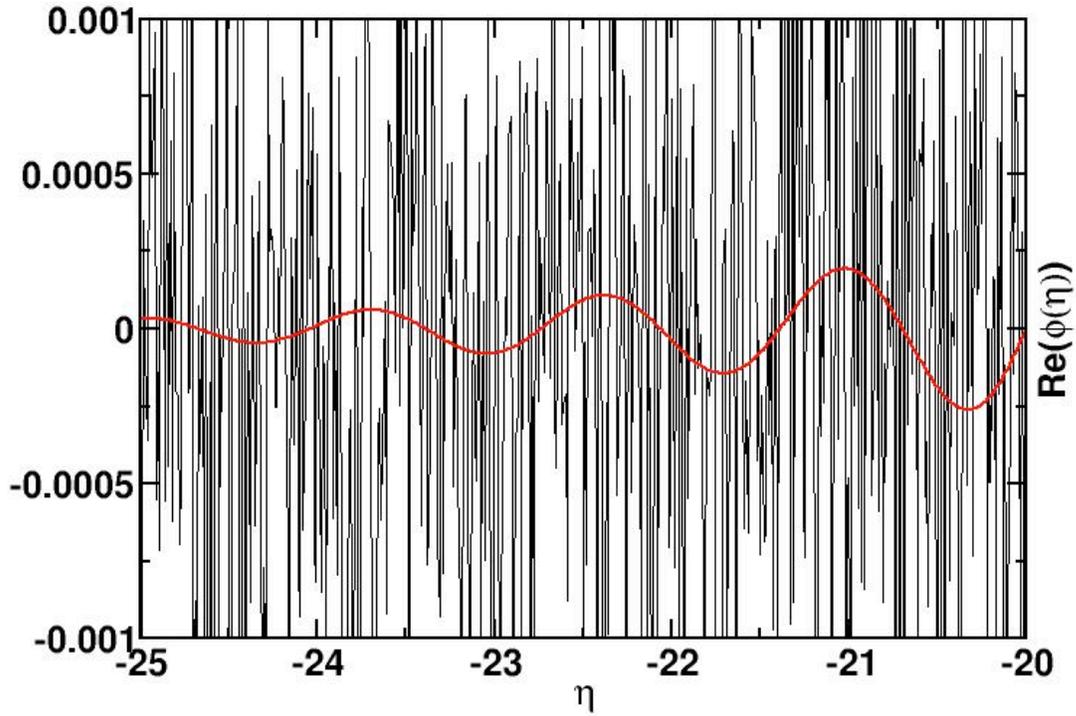

Fig. 7: Expanded detail of Re($\phi(\eta)$) from Fig. (5) for $x_0 = 0.50$ for large, negative $\eta$ values, $\varepsilon = 0.40$ (red) vs corresponding direct Monte Carlo results (black).

Finally, Figs. (8) and (9) display the real and imaginary portions of $\phi(\eta)$ computed for a fixed value of $x_0$ (0.5) and varying values of $\sigma$ (1.00 = black, 2.00 = red, 3.00 = green) using the present approach. All results utilize $10^6$ Monte Carlo points for the evaluation of the necessary $\langle x \rangle_\lambda$ results.

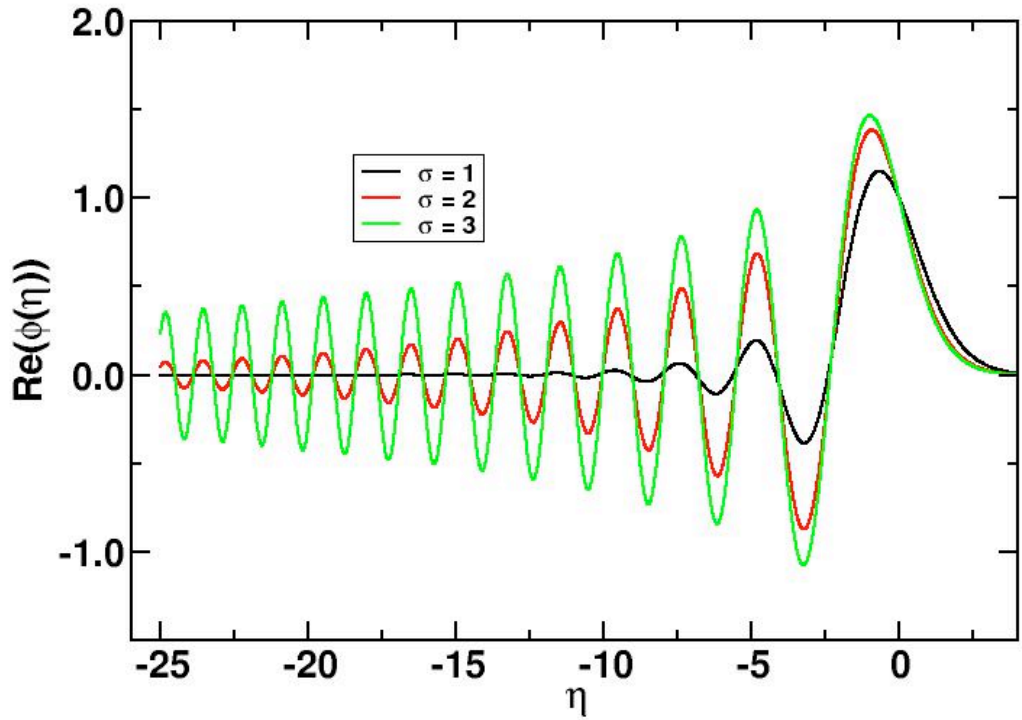

Fig. 8: Plots of the real parts of $\phi(\eta)$ for Gauss/Airy model computed via the Kirkwood approach. Results shown correspond to $x_0 = 0.5$ and were computed using 100 loops of $10^4$ points for $\sigma = 1.00$ (black), 2.00 (red) and 3.00 (green).

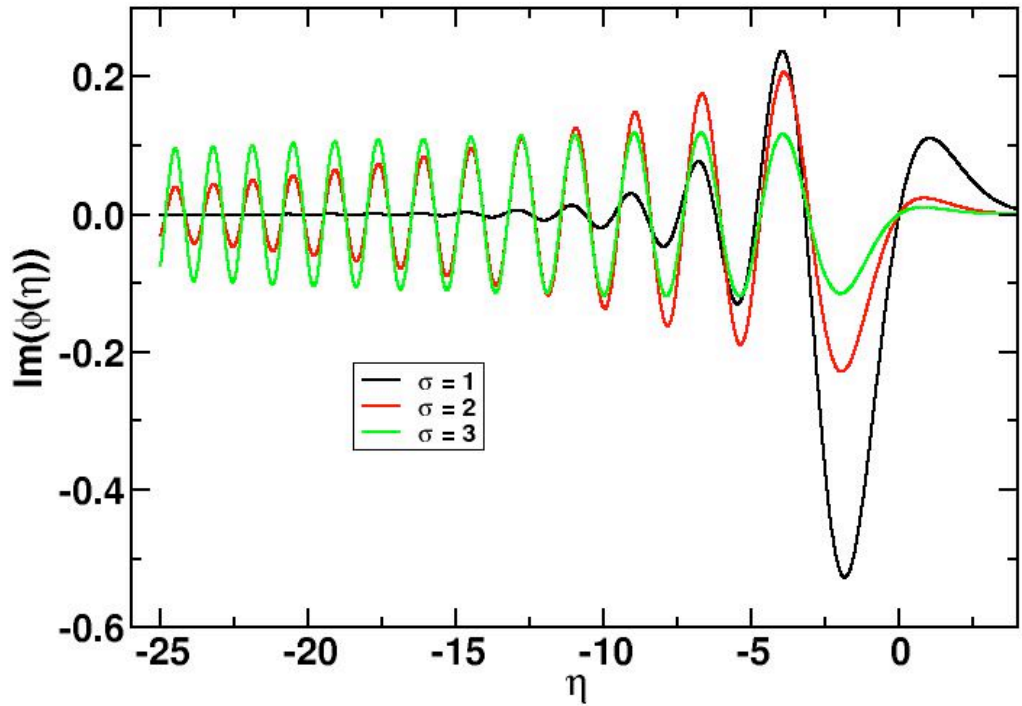

Fig. 9: As in Fig. (8), but for the imaginary parts of ϕ(η).

***V. Summary and Future Directions:***  In the present work we have explored the problem of calculating generalized averages.  We have presented an approach that combines thermodynamic integration and Stationary Phase Monte Carlo techniques to cope with the risk-sensitive and rare-event sampling issues involved and have explored its application to a prototypical class of problems.

We close by noting that the methods developed in the present work would appear to offer a potential tool for the study of real-time quantum dynamics.  In particular, we note that a generic coordinate-space, equilibrium time correlation function, $G_{AB}(t)$, can be expressed as

$$G_{AB}(t) = \frac{\int dx\, dx'\, \rho(x) P(x \to x', t) A(x) B(x')}{\int dx\, dx'\, \rho(x) P(x \to x', t)},$$

(5.1)

where ρ(x) is proportional to the probability of x and P(x→x',t) is the conditional probability density that the system that starts at x at time zero arrives at x' a time t later. If one can sample these "initial" and "final" conditions, we can approximate $G_{AB}(t)$ as

$$G_{AB}(t) \approx \frac{1}{N} \sum_{n=1}^{N} A(x_n) B(x_n'),$$

(5.2)

where the points $\{x_n\}$ are a random sample of ρ(x) and the points $\{x_n'\}$ are a random sample of the conditional probability $P(x_n \rightarrow x', t)$.

Generating a sampling of the initial positions, $\{x_n\}$, is a standard equilibrium problem, one for which well established classical[2] and quantum-mechanical approaches[3-5] exist. Techniques for sampling the conditional probability involved are well established for classical systems, but generally lacking for quantum-mechanical ones.

The minimal information needed to sample the conditional probability, P(x→x',t), with respect to the final position, x', is knowledge of ratios of the form

$$|R|^2 = \frac{P(x \rightarrow x'', t)}{P(x \rightarrow x', t)}.$$

(5.3)

As noted previously,[5] R for finite temperature quantum-mechanical problem can be written in Kirkwood form. The combination of these Kirkwood expressions for R, the current INS/SPMC approach for their evaluation, and previously developed penalty Monte Carlo methods[17] would appear to offer a possible approach to the conditional, quantum-mechanical sampling problem. Time will tell if this is so.

*Acknowledgments:* The authors wish to acknowledge support from the National Science Foundation award DMS-1317199 and from the DARPA EQUiPS award W911NF-15-2-0122.